\documentclass[twocolumn,aps,10pt,showkeys,showpacs,preprintnumbers,prd,superscriptaddress,nofootinbib]{revtex4-1}
\usepackage{graphicx}	
\usepackage{amssymb}
\usepackage{times}
\usepackage{dcolumn}
\usepackage{float}
\usepackage[compat=1.1.0]{tikz-feynman}
\usepackage{mathtools}
\usepackage{amsmath}
\usepackage{xcolor}
\usepackage{color}
\usepackage{subcaption}
\usepackage{booktabs}
\usepackage{comment}
\usepackage{tikz}
\usepackage{pgfplots}
\usepackage{tikz-feynman}
\usepackage[normalem]{ulem}
\usepackage{theorem}
\usepackage{rotating, bm, array}
\usepackage[pagebackref=false, colorlinks=true]{hyperref}
\hypersetup{linkcolor=blue, citecolor=blue,urlcolor=blue}

\begin{document}
\title{Mass and Decay Properties of Toponium Using SUSY QM Factorization Method}

\author{Bhavishya Tepan}
\email{bhavishyatepan269@gmail.com}
\author{Manan Shah}
\email{mnshah09@gmail.com}
\author{P C Vinodkumar}
\email{p.c.vinodkumar@gmail.com}
\affiliation{P. D. Patel Institute for Applied Sciences, Charusat University, Anand-388421, Gujarat, India}

\date{\today}
\begin{abstract}

The Supersymmetric quantum mechanics (SUSY QM) factorization method is applied to study the Cornell potential in the context of toponium $(t\bar{t})$, the heaviest quarkonium system. The method is succesfully applied to determine the mass, binding energy, and spin-dependent hyperfine splitting, giving pseudoscalar $(\eta_t)$ and vector $(\Theta_t)$ state masses of 344.114 GeV and 344.141 GeV, respectively. The decay properties of both the pseudoscalar and vector states are also investigated through their respective decay widths, highlighting the crucial role of the bound-state wavefunction. Among the decay channels, the di-gluonic $(\eta_t\rightarrow gg)$ mode is found to be dominant, with an estimated decay width of 2.57 MeV. In addition, the spatial characteristics of toponium are examined by evaluating the mean radius, root-mean-square (RMS) radius, and most probable radius, all of which are found to lie within the range of $0.015\text{--}0.025\,\mathrm{fm}$. \\

\end{abstract}
\maketitle

\section{Introduction} 
Toponium represents the quasi bound state of a top quark and its corresponding antiquark. The top quark is the heaviest quark flavour discovered so far with the mass of $m_t=172.52\pm0.33$ GeV, the top quark generally decays too quickly—in roughly $3.29\times10^{-25}$ seconds \cite{PhysRevD.110.030001,Abazov_2012}. However, when these pairs are generated with very little relative motion, quantum effects such as entanglement allow them to interact through gluon exchange for a minimal duration before they decay \cite{Varma_2024,francener2025investigatingexclusivetoponiumproduction}. This state, technically designated as $\eta_t/\Theta_t$, is remarkably small, possessing a Bohr radius of approximately $8\times10^{-18}$ meters \cite{Top_BS}. Unlike lighter versions of these pairings, such as charmonium or bottomonium, toponium is so heavy that it decays predominantly through weak interaction rather than through electromagnetic or strong force processes \cite{Bigi:1980az}. For decades, the scientific community has doubted whether the toponium state would be possible and would be detected at the Large Hadron Collider (LHC) because the events are rare and difficult to distinguish from background noise. This belief changed in mid-2025 when the ATLAS and CMS collaborations identified a clear surplus of top-quark pairs at the minimum energy threshold. They have now confirmed the state's existence with a statistical significance reaching more than $8\,\sigma$, effectively proving that the top and anti-top quarks do indeed pair up briefly \cite{2023,2025,Collaboration_2025,atlascollaboration2026observationcrosssectionenhancementnear}. In this context, future high-precision collider experiments—including the Circular Electron Positron Collider, the Future Circular Collider, and proposed muon colliders—are expected to focus their study on heavy flavour hadrons containing $t / \bar t$ quark in them. \cite{thecepcstudygroup2018cepcconceptualdesignreport,thecepcstudygroup2025cepctechnicaldesignreport,agapov2022futurecircularleptoncollider,accettura2025interimreportinternationalmuon,Han_2025}. Such facilities would enable extremely accurate measurements of top quark pair production cross-sections, decay channels, and threshold behavior, significantly improving our understanding of the underlying dynamics. Toponium dynamics offers a uniquely sensitive probe of the Standard Model (SM), primarily due to its strong dependence on the top–Higgs Yukawa coupling, electroweak symmetry breaking, and the structure of the quantum vacuum \cite{PhysRevD.43.264,SEHGAL1981417,Voloshin:1978hc}. As the heaviest known quark, the top quark couples most strongly to the Higgs field, making any bound-state or near-threshold dynamics of top–antitop pairs especially informative for testing the consistency of the Standard Model (SM). Studying toponium provides a unique opportunity to probe physics beyond the Standard Model, such as effects of quantum gravity or the presence of extra spatial dimensions \cite{Dom_nech_2021,Giudice_1999}. As a result, toponium serves as a valuable system for exploring possible extensions to our current understanding of fundamental interactions.\\
Modern theoretical studies using the Salpeter equation and the Cornell potential predict that the ground (1S) state has a mass value of 343.62 GeV \cite{PhysRevD.111.096016}. Calculations using a pure Coulomb potential yield a slightly lower mass of 343.59 GeV for the same 1S state \cite{PhysRevD.111.096016}. Because the top quark is so heavy ($m_t$ =172.7 GeV), relativistic corrections are considered negligible, and the mass splittings between different spin states are effectively zero. In Ref. \cite{https://doi.org/10.1155/ahep/9454944}, the authors utilize a non-relativistic quark potential model (NRQPM), specifically a Cornell potential incorporating spin-dependent interactions, and reported the $1^1S_0$  state mass at 342.867 GeV. Another paper \cite{Top_BS} presents a specialized approach for discovering the vector toponium state ($J_t$) and determining the scheme-independent 1S top quark mass with high accuracy and by examining the ratio of cross sections ($R_b$) in $e^+ e^-$ collisions around the production threshold of approximately 341 GeV.

Previous studies have indicated that, in addition to the dominant decay channel $\eta_t \to W^+ b\, W^- \bar{b}$, there exist conventional signatures arising from the gluonic decay $\eta_t \to gg$, with a relatively small decay width of about $1.99\,\text{MeV}$ and a branching ratio of $6.63 \times 10^{-4}$ \cite{Yang__2026}. Furthermore, analyses have also estimated other important decay modes, including the leptonic vector decay channel, $\Theta_t \to \ell^+ \ell^-$, with a decay width of approximately $6.09\,\text{keV}$ \cite{PhysRevD.111.096016}, and the di-gamma decay, $\eta_t \to \gamma\gamma$, which is predicted to have a decay width of around $9.32\,\text{keV}$ \cite{Yang__2026}. The other vector decay channel such as $\gamma\gamma\gamma$ and $ggg$ are found to have a decay width of 15 eV and 52.7 eV, respectively \cite{jiang2024studytoponiumspectrumassociated}. Similarly, the results also indicate that the $\eta_t \rightarrow Z^{0}H$ channel is the dominant electroweak mode, with a partial width of $537.43~\text{keV}$ and a branching ratio of $1.79 \times 10^{-4}$ \cite{Yang__2026}. In contrast, the $\eta_t \rightarrow Z^0\gamma$ channel is one of the most suppressed, with a partial decay width of only $2.01~\text{keV}$ and a branching ratio of $6.71 \times 10^{-7}$ \cite{Yang__2026}. The $\eta_t \rightarrow W^{+}W^{-}$ decay yields a partial width of $97.45~\text{keV}$ and branching ratio of $3.25 \times 10^{-5}$, with the $\eta_t \rightarrow Z^{0}Z^{0}$ and $\eta_t \rightarrow \gamma\gamma$ modes each carrying branching ratios of $3.11 \times 10^{-6}$ \cite{Yang__2026}. The vector bosonic decay results predict the $W^{+}W^{-}$ channel to be the dominant mode, with a branching ratio of $9.96 \times 10^{-4}$, while the $Z^0Z^0$ channel is among the most suppressed, with a branching ratio of $9.29 \times 10^{-6}$ \cite{Bai_2026}. The fermionic pseudoscalar decay modes are found to be strongly suppressed, being proportional to the square of the fermion mass, with the $b\bar{b}$ channel carrying a branching ratio of only $7.36 \times 10^{-8}$. In contrast, for the vector state, the $b\bar{b}$ channel is predicted to be one of the dominant fermionic decay modes, with a branching ratio of $8.54 \times 10^{-4}$ \cite{Bai_2026}. The characteristic size of the ground-state toponium system has also been shown to be extremely small, with the 1S state rms radius estimated to be $0.018\,\text{fm}$ \cite{https://doi.org/10.1155/ahep/9454944}. This remarkably compact scale reflects the large mass of the top quark emphasizing the highly localized nature of the toponium bound state, whose mass according to the recent CMS observations near the top quark pair production threshold suggest to lie approximately in the range of \(338\text{--}350\,\text{GeV}\) \cite{Collaboration_2025}, consistent with the interpretation of a near-threshold \(t\bar{t}\) quasi-bound state .

To understand the complex forces that govern such bound quantum systems, physicists often use Supersymmetric Quantum Mechanics (SUSY QM) as a powerful analytical framework \cite{parmar2002supersymmetric,schwabl2007quantum}. At its core, SUSY QM provides a structured way to simplify the Hamiltonian of a system by factorizing it into more elementary operators known as supercharges. This factorization connects two different but related quantum systems, called partner potentials \cite{I1_Cooper_1995,tong2007susyqm,khare1997}. These partner systems share nearly identical energy spectra, differing only in their ground state, which makes the method particularly useful for solving otherwise difficult problems \cite{ias_akshayku_2003,Napsuciale_2021}.
A particularly important application of this framework lies in the study of the Cornell potential, which is widely used to model the interaction between quarks inside mesons \cite{Alkathiri_2024}. Using methods like the Supersymmetric Expansion Algorithm (SEA), physicists can obtain analytical solutions that reveal important properties of heavy quarkonium systems such as charmonium and bottomonium, and even predict new, yet-unobserved states
\cite{Napsuciale_2025}.

\section{SUSY QM}

In this section, we highlight the general mathematical framework of Supersymmetric Quantum Mechanics (SUSY QM). The starting point is the factorization of the Hamiltonian in terms of ladder operators. A quantum system with a Hamiltonian
\begin{equation}
    H_0 = -\frac{1}{2} \frac{d^2}{dx^2} + V_0(x),
\end{equation}
can be factorised via ladder operators as
\begin{equation}
    H_\ell=a^\dagger_\ell a_\ell + C(\ell,\lambda),
\end{equation}
where
\begin{equation}
    a_\ell=\frac{d}{dx}+W_\ell ,\qquad a^\dagger_\ell=-\frac{d}{dx}+W_\ell.
\end{equation}

Thus, the superpotential $W_\ell$ must satisfy the condition,
\begin{equation}
    W_\ell^2(x) - W_\ell'(x) + C(\ell) = v_\ell(x).
\end{equation}

This gives us a first order non linear differential equation known as Ricatti equation, which after solving yields the partner Hamiltonian as
\begin{equation}
    \tilde{H}_\ell = a_\ell a_\ell^\dagger  + C(\ell, \delta) = -\frac{d^2}{dx^2} + \tilde{v}_\ell(x).
\end{equation}

To organize SUSY QM more formally, we define a two-component Hamiltonian:
\begin{equation}
\mathcal{H} =
\begin{pmatrix}
a_\ell a_\ell^\dagger & 0 \\
0 & a_\ell^\dagger a_\ell
\end{pmatrix},
\end{equation}
which can be written in terms of the supercharges
\begin{equation}
Q_1=\begin{pmatrix}
    0 & -ia_\ell\\
    ia_\ell^{\dagger} & 0
\end{pmatrix},\qquad 
Q_2=\begin{pmatrix}
    0 & a_\ell\\
    a_\ell^{\dagger} & 0
\end{pmatrix}.
\end{equation}

These two operators satisfy \(\{Q_i , Q_j\} = 2\delta_{ij}H\), \([Q_i , H] = 0\) which represents that both operators are true constants of motion.

With the general SUSY QM framework established, we now move on to applying it to a specific potential suited for quarkonium systems. The choice of potential is important, as it should capture the key features of the quark-antiquark interaction. In this work, we use the Cornell potential. It is also known as the linear plus Coulomb potential and is an effective method to explain the confinement of quarks in quantum chromodynamics (QCD). The potential has the form:
\begin{equation}
V(r)=-\frac{4}{3}\frac{\alpha_s}{r}+\sigma r
\end{equation}

The potential consists of two parts. The first one, 
\(V(r)=-\frac{4}{3}\frac{\alpha_s}{r}\)
at short distances, is known as the colour Coulombic part of the potential, since it has the same form as the well-known Coulombic potential
\(-\frac{\alpha_s}{r}\) where $\alpha_s$ is the strong coupling constant.
The second term of the potential, 
\(\sigma r\) is the linear confinement term, where $\sigma$ is the string tension. The dimensionless radial Schrodinger equation can be reduced to
\begin{equation}\label{radialsch}
\left[-\frac{d^2u(x,\lambda)}{dx^2}+v_0(x,\lambda)\right]u(x,\lambda)=\epsilon_0(\lambda)u(x,\lambda),
\end{equation}
where, the dimesionless effective potential and energy is given by
\begin{equation}
v_0(x,\lambda)=\frac{\ell(\ell+1)}{x^2}-\frac{2}{x}+\lambda x, \,\,\,\,\,\,
 \epsilon_0(\lambda)=\frac{E}{\mu \alpha^2},
\end{equation}
with $\lambda$ being the normalized string tension parameter defined as $\lambda=\frac{2\sigma}{\mu^2\alpha^3}$ and $x=r/a_0$. Here, \(a_0=\frac{\hbar}{\mu c\alpha}\)\footnote{For a general Coulomb-like potential $V(r)=-\kappa/r$, minimizing the total energy $E(r)=\hbar^{2}/(2\mu r^{2})-\kappa/r$ with respect to r gives the characteristic Bohr radius $a_{0}=\hbar^{2}/(\mu\kappa)$. In the case of heavy quarkonia, the potential is $V(r)=-C_{F}\alpha_{s}\hbar c/r$, so that $\kappa=C_{F}\alpha_{s}\hbar c$, giving $a_{0}=\hbar/(\mu c\,C_{F}\alpha_{s})$.} denotes the Bohr radius which serves as the characteristic scale that removes dimensions from the problem, transforming the Schrödinger equation into a dimensionless form and simplifying both analytical and numerical calculations, and $\alpha=\frac{4}{3}\alpha_s$ is the effective coupling constant \cite{Napsuciale_2025}.

The toponium system is analyzed within the above described framework. The large top quark mass ensures that the interaction is dominated by short-distance dynamics. For simplicity, the strong coupling constant ($\alpha_s$) is evaluated at a fixed scale, $\mu = m_t$.
\begin{equation}
\alpha_s(\mu) = \frac{12\pi}{33 - 2N_f} \, \frac{1}{\log\left(e + \frac{\mu^{\,2}}{\Lambda^2}\right)},
\end{equation}
where $e = 2.71828$, $N_f = 5$ corresponds to the number of active quark flavors relevant for the $t\bar{t}$ system, and $\Lambda = 0.10 \, \text{GeV}$ is the QCD scale parameter. Using these inputs together with the top quark mass $m_t = 172.52 \, \text{GeV}$ \cite{PhysRevD.110.030001}, we obtain ${\alpha_s(m_t) \approx 0.11}$, which sets the effective strength of the interaction in our analysis.

The solution of equation (\ref{radialsch}) is obtained as per the algorithm discussed in \cite{Napsuciale_2025}. Here, the linear confining interaction is treated as a perturbation around the exactly solvable Coulomb problem, and the superpotential is expanded as a power series in terms of the normalized string tension $(\lambda)$. Performing the expansion, we obtain the square of the wavefunction at the origin as
\begin{equation}
\lvert{\Psi(0)\rvert}^2=\frac{1}{\pi a_0^3}\left(1+\frac{3}{2}\lambda-\frac{31}{16}\lambda^2\right).
\end{equation}

\subsection{Mass and Energy}
With the interaction strength specified, the mass of a quarkonium system is determined by the sum of the constituent quark masses along with the contribution from the binding energy. The binding energy arises from the interaction between the quark and antiquark and reflects the stability of the bound state. In general, it can be obtained from the energy eigenvalues of the system. Using the derived results, the binding energy is found to be $\mathbf{E_B = -0.9056 \, \text{GeV}}$. Using this result, the corresponding mass of the toponium system is calculated by adding the individual top quark masses with the binding energy,
\begin{equation}
m_{t\bar t}=2m_{t}+E_B,
\end{equation}
providing the spin average mass of $t \bar t$ (toponium) to be about ${m_{t \bar t}^{(1S)}=344.134\, \text{GeV}}$. Correspondingly, we find the hyperfine splitting $\Delta E_{hf}$ as \cite{saha_bulk_viscosity},
\begin{equation}
    \Delta E_{hf}=\frac{32\pi \alpha_s}{9m_Q^2}|\Psi(0)|^2\langle \mathbf{S_1\cdot S_2}\rangle.
\end{equation}
Now,
\[\langle\mathbf{S}_1 \cdot \mathbf{S}_2 \rangle
=
\begin{cases}
-3/4 & S=0 \\
+1/4 & S=1
\end{cases}.
\]

Thus, we get the following pseudoscalar and vector state masses for toponium as Table \ref{tab:placeholder}.

\begin{table}[H]
    \centering
    \setlength{\tabcolsep}{4pt}
\begin{tabular}{lccccc}
\hline
    State & [Our] &
    Ref.\cite{Najjar2026} & Ref.\cite{https://doi.org/10.1155/ahep/9454944} & Ref.\cite{PhysRevD.111.096016} & Ref.\cite{jiang2024studytoponiumspectrumassociated}\\
    \hline
    \hline
        $\eta_t$ (GeV) & 344.114  &  343.53 & 342.87 & $\sim$343.62 & 341.27\\
       $\Theta_t$ (GeV) & 344.141 & 343.59 & 342.91 & $\sim$343.62 & 341.65\\
    \hline
\end{tabular}
    \caption{Pseudoscalar $(0^{-+})$ and Vector $(1^{--}$) state mass of Ground state (1S) of toponium. }
    \label{tab:placeholder}
\end{table}
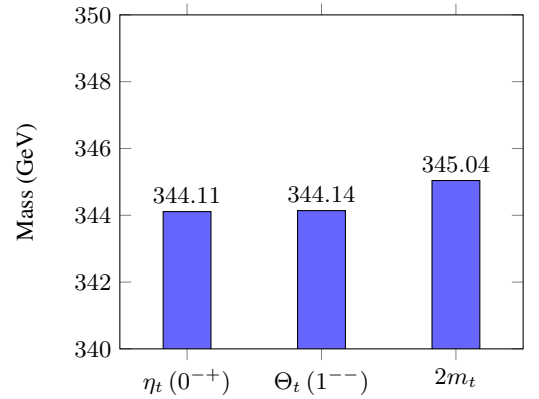
\begin{figure}[H]
\centering
\begin{tikzpicture}
\begin{axis}[
    ybar,
    bar width=18pt,
    ymin=340, ymax=350,
    ylabel={Mass (GeV)},
    symbolic x coords={$\eta_t\,(0^{-+})$, $\Theta_t\,(1^{--})$, $2m_t$},
    xtick=data,
    nodes near coords,
    nodes near coords align={vertical},
    enlarge x limits=0.25,
    width=0.8\columnwidth,
    height=6cm
]
\addplot[
    fill=blue!60,
    draw=black] coordinates {
    ($\eta_t\,(0^{-+})$, 344.11)
    ($\Theta_t\,(1^{--})$, 344.14)
    ($2m_t$, 345.04)
};

\end{axis}
\end{tikzpicture}
\caption{Comparison of pseudoscalar and vector toponium masses with twice the top quark mass.}
\end{figure}
\subsection{Radii}
The spatial extent of the bound state can be analyzed through the radial probability density, which provides information about the distribution of the quark-antiquark separation. The radial probability density upto $\mathcal{O}(\lambda^2)$ is calculated as  
\begin{equation}
u^2(x) = x^2 e^{-2x} \left[ 1 - \frac{x^2}{2}\lambda + \left(\frac{x^2}{4} + \frac{x^3}{12} + \frac{x^4}{8}\right)\lambda^2 \right].
\end{equation}

This expression can be used to calculate significant physical values such as the mean radius, root mean square (RMS) radius, and most probable radius. The mean radius represents the average separation between quarks and antiquarks, whereas the RMS radius measures the spatial spread of the bound state. These quantities are defined using the spatial coordinate's expectation values as  
\begin{equation}
\langle x \rangle = \frac{\int_0^\infty x\, u^2(x)\, dx}{\int_0^\infty u^2(x)\, dx}, \qquad
\langle x^2 \rangle = \frac{\int_0^\infty x^2\, u^2(x)\, dx}{\int_0^\infty u^2(x)\, dx},
\end{equation}
\[\text{with}\,\,x_{\text{rms}} = \sqrt{\langle x^2 \rangle}.\]
 where, $\langle r\rangle=a_0\langle x \rangle$ and $r_{rms}=a_0x_{rms}$ gives the following results, $\mathbf{0.02304\,fm}$ and $\mathbf{0.02657\,fm}$ respectively.
 
In addition to these averaged quantities, the most probable radius corresponds to the point at which the radial probability density attains its maximum value, representing the most likely separation between the quark and antiquark. A plot of $u^2(r)$ as a function of $r$ shows this behaviour, with the peak of the curve representing the most probable radius as $\mathbf{r_{mp}=0.01544\,\,fm}$ . 

\begin{figure}[h]
\centering
\begin{tikzpicture}
\begin{axis}[
    width=8cm,
    height=7cm,
    xlabel={$r\;(\mathrm{fm})$},
    ylabel={$P(r)=|u(r)|^2\;(\text{fm}^{-1})$},
    xmin=0, xmax=0.16,
    ymin=0, ymax=6000,
    xtick={0,0.02,0.04,0.06,0.08,0.10,0.12,0.14},
    ytick={0,1000,2000,3000,4000,5000,6000},
    xticklabels={0,0.02,0.04,0.06,0.08,0.10,0.12,0.14,0.16},
    yticklabels={0,1000,2000,3000,4000,5000},
    grid=major,
    samples=200,
    domain=0:0.16,
    thick
]
\pgfmathsetmacro{\lam}{0.016}
\pgfmathsetmacro{\mu}{172.52/2}
\pgfmathsetmacro{\alpha}{4*0.11/3}
\pgfmathsetmacro{\a}{1/(\mu*\alpha)}
\pgfmathsetmacro{\hbarc}{0.1973269804}
\pgfmathsetmacro{\Nnorm}{
sqrt((1/(pi*(\a)^3))*(1+3*\lam/2-31*(\lam^2)/16))
}
\addplot[
    blue,
    thick
]
{
    (\Nnorm)^2/(\a*\hbarc) *
    (x/(\a*\hbarc))^2 *
    exp(-2*(x/(\a*\hbarc))) *
    (1-\lam*(x/(\a*\hbarc))^2/2)
};
\end{axis}
\end{tikzpicture}
\caption{Radial probability distribution of toponium.}
\end{figure}
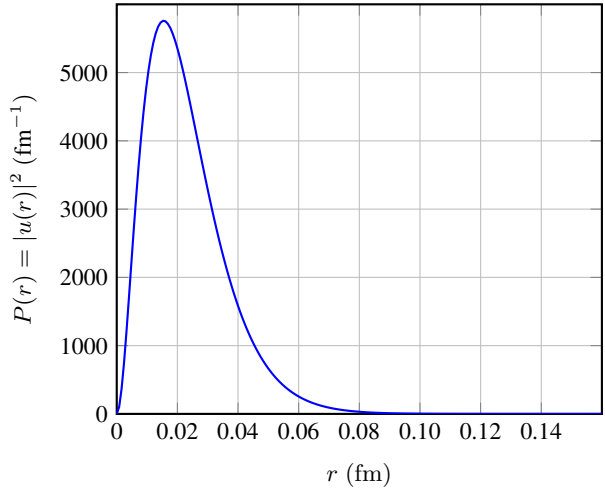

The hyperfine splittings estimated using the most probable radius, $r_{mp}$, and the root mean square radius, $r_{rms}$, are found to be $3.72$ MeV and $0.88$ MeV, respectively. These values are smaller than the hyperfine splitting $\Delta E_{hf}=27.20$ MeV obtained from the wavefunction at the origin, indicating a reduction in the spin-spin interaction strength with increasing interquark separation.
\subsection{Decay Widths}
Another important feature of heavy quark-antiquark systems is their annihilation decay width, which describes the pair's likelihood of annihilating into other particles. In these decay processes, the quark–antiquark pair annihilates into gauge bosons, such as gluons, photons, or electroweak bosons, or into fermion–antifermion pairs through the corresponding interaction channels. In this work, we employ the various decay coupling constants, and kinematic factors as defined in previous works \cite{jiang2024studytoponiumspectrumassociated,PhysRevD.37.3210,Yang__2026,Fabiano1994,PhysRevD.35.3366,Bai_2026}. The electroweak couplings are expressed in terms of $\alpha_{\mathrm{em}}$ and the weak mixing angle $\theta_W$, such that
\begin{equation}
\alpha_Z = \frac{\alpha_{\mathrm{em}}}{\sin^2\theta_W \cos^2\theta_W}, \quad
\alpha_W = \frac{\alpha_{\mathrm{em}}}{\sin^2\theta_W}.
\end{equation}
Furthermore, the dimensionless mass ratio is defined as
\begin{equation}
R_i = \frac{m_i^2}{M_{\eta_t}^2},
\end{equation}
where $m_i$ is the mass of the final-state particle. The color factor $N_f$ in fermionic decay takes the value 1 for leptons and 3 for quarks, reflecting their respective color degrees of freedom. Finally, the kinematic function
\begin{equation}
\lambda(a,b,c) = a^2 + b^2 + c^2 - 2ab - 2bc - 2ca,
\end{equation}
known as the Källén function, is used to describe phase-space factors in two-body decays
\cite{Kallen:1964lxa}.
\subsubsection{Gluonic Decay Width}

Gluonic decay processes are governed by Quantum Chromodynamics (QCD) and typically provide the dominant contribution in regimes where electromagnetic decay channels are suppressed. These transitions arise from the annihilation of the quark--antiquark pair into gluons. Accordingly, the di-gluonic decay of the pseudoscalar state, ($\eta_t$) and the tri-gluonic decay of the vector state, ($\Theta_t$) are given by \cite{jiang2024studytoponiumspectrumassociated,Fabiano1994}.
\begin{equation}
\Gamma(\eta_t\,\rightarrow gg)=\frac{32\pi \alpha_s^2 }{3M_{\eta_t}^2}
|\psi(0)|^2\left(1+\frac{3.99}{\pi}\alpha_s\right)
\end{equation}
\begin{equation}
\Gamma(\Theta_t\,\rightarrow ggg)=\frac{160(\pi^2-9) \alpha_s^3 }{81M_{\Theta_t}^2}
|\psi(0)|^2\left(1-\frac{6.06}{\pi}\alpha_s\right)
\end{equation}
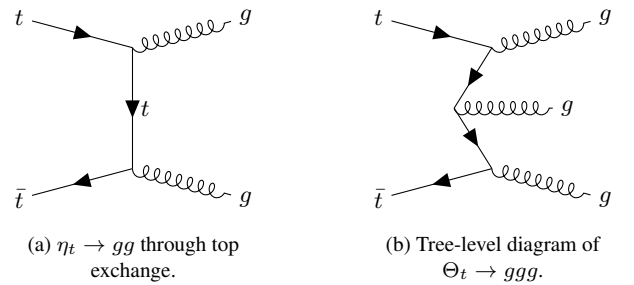
\begin{figure}[h]
\centering
\begin{subfigure}[b]{0.45\columnwidth}
\centering
\begin{tikzpicture}
\begin{feynman}

\vertex (i1) at (0.5,1.2) {\(t\)};
\vertex (i2) at (0.5,-1.2) {\(\bar t\)};

\vertex (v1) at (2,0.8);
\vertex (v2) at (2,-0.8);

\vertex (o1) at (3.5,1.2) {\(g\)};
\vertex (o2) at (3.5,-1.2) {\(g\)};

\diagram*{
(i1) -- [fermion] (v1)
      -- [fermion, edge label=\(t\)] (v2)
      -- [fermion] (i2),

(v1) -- [gluon] (o1),
(v2) -- [gluon] (o2),
};

\end{feynman}
\end{tikzpicture}
\caption*{\footnotesize(a) \(\eta_t \rightarrow gg\) through top exchange.}
\end{subfigure}
\hfill
\begin{subfigure}[b]{0.45\columnwidth}
\centering
\begin{tikzpicture}
\begin{feynman}

\vertex (i1) at (0.5,1.2) {\(t\)};
\vertex (i2) at (0.5,-1.2) {\(\bar t\)};

\vertex (v1) at (2,0.8);
\vertex (v2) at (1.5,0);
\vertex (v4) at (2,-0.8);

\vertex (o1) at (3.5,1.2) {\(g\)};
\vertex (o2) at (3.5,-1.2) {\(g\)};
\vertex (o3) at (3,0) {$g$};
\diagram*{
(i1) -- [fermion] (v1)
      -- [fermion] (v2)
      -- [fermion] (v4)
      -- [fermion] (i2),

(v2) -- [gluon] (o3),
(v1) -- [gluon] (o1),
(v4) -- [gluon] (o2),
};

\end{feynman}
\end{tikzpicture}
\caption*{\footnotesize(b) Tree-level diagram of \(\Theta_t\to ggg\).}
\end{subfigure}
\caption{Dominant hadronic decay channels of pseudoscalar and vector toponium.}
\end{figure}

\subsubsection{Photonic Decay Width}

Even though electromagnetic decay channels, such as photonic decays, are typically suppressed, they remain important because of their clean experimental signatures. Even when such decays proceed through higher-order or indirect mechanisms, the absence of strong interaction backgrounds makes them particularly valuable for precision measurements and for testing theoretical predictions. For quarkonia, these decay widths are computed as \cite{PhysRevD.37.3210,jiang2024studytoponiumspectrumassociated}.

\begin{equation}
\Gamma(\eta_t\,\rightarrow \gamma \gamma)=\frac{48\pi \alpha_{em}^2 e_Q^4}{M_{\eta_t}^2}
|\psi(0)|^2\left(1-\frac{3.4}{\pi}\alpha_s\right)
\end{equation}
\begin{equation}
\Gamma(\Theta_t\,\rightarrow \gamma \gamma \gamma)=\frac{64(\pi^2-9) \alpha_{em}^3 e_Q^6}{3M_{\Theta_t}^2}
|\psi(0)|^2\left(1-\frac{12.6}{\pi}\alpha_s\right)
\end{equation}
\begin{figure}[h]
\centering

\begin{subfigure}[b]{0.45\columnwidth}
\centering
\begin{tikzpicture}
\begin{feynman}

\vertex (i1) at (0,1.2) {\(t\)};
\vertex (i2) at (0,-1.2) {\(\bar t\)};

\vertex (v1) at (1.5,0.8);
\vertex (v2) at (1.5,-0.8);

\vertex (o1) at (3.5,1.2) {\(\gamma\)};
\vertex (o2) at (3.5,-1.2) {\(\gamma\)};

\diagram*{
(i1) -- [fermion] (v1)
      -- [fermion, edge label=\(t\)] (v2)
      -- [fermion] (i2),

(v1) -- [boson] (o1),
(v2) -- [boson] (o2),
};

\end{feynman}
\end{tikzpicture}
\caption*{\footnotesize(a) \(\eta_t \rightarrow \gamma\gamma\) through top exchange.}
\end{subfigure}
\hfill
\begin{subfigure}{0.45\columnwidth}
\centering
\begin{tikzpicture}

\begin{feynman}

\begin{tikzpicture}

\begin{feynman}


\vertex (tbar) at (0,1.2+1.85) {\(t\)};

\vertex (t)    at (0,-1.2+1.85) {\(\bar t\)};

\vertex (v1) at (1.5,0.8+1.85);

\vertex (v2) at (1.5,-0.8+1.85);

\vertex (v3) at (2,-1.4+1.85);

\vertex (g1) at (3.5,1.2+1.85) {\(\gamma\)};

\vertex (g2) at (3.5,-0.8+1.85) {\(\gamma\)};

\vertex (g3) at (3.5,-1.45+1.85) {\(\gamma\)};

\diagram*{


(tbar) -- [fermion] (v1),

(t)    -- [anti fermion] (v2),


(v1) -- [fermion] (v2),

(v2) -- [fermion] (v3),


(v1) -- [photon] (g1),

(v2) -- [photon] (g2),

(v3) -- [photon] (g3),

};

\end{feynman}

\end{tikzpicture}
\end{feynman}

\end{tikzpicture}
\caption*{\footnotesize(b) Tree-level diagram of $\Theta_t \to \gamma\gamma\gamma$}
\end{subfigure}

\caption{Electromagnetic decay channels of pseudoscalar and vector toponium.}
\end{figure}
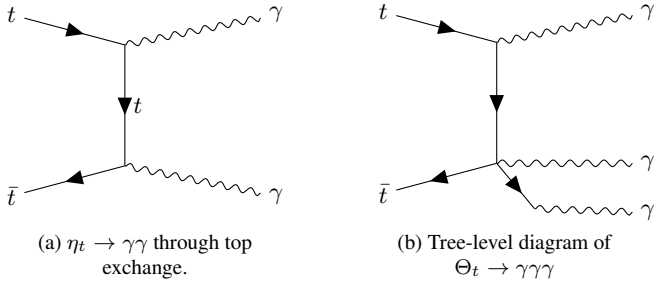

\subsubsection{Fermionic Decay Width}
The fermionic decay modes which involves the production of fermion–antifermion pairs, are particularly sensitive to mass-dependent couplings and thus offer insight into the interplay between bound-state dynamics and electroweak interactions \cite{Bagnaschi_2026}. The SUSY QM approach, through its shape-invariant construction and analytic control over the spectrum, provides the bound-state properties required to systematically study the impact of mass effects on decay observables such as \cite{Yang__2026,PhysRevD.35.3366}.
\begin{equation}
\Gamma(\eta_t\,\rightarrow f\bar{f}) = N_f \frac{3\pi\alpha_Z^2}{8M_{\eta_{t}}^2} \frac{R_f}{R_Z^2}
\lambda^{1/2}(1, R_f, R_f)|\psi(0)|^2
\end{equation}

{\small
\begin{align}
\Gamma(\Theta \to f\bar{f}) =
\frac{16\pi\alpha_{em}^{2} N_f \beta_f}{M_{\Theta_t}^{2}}
\Bigg[
(1+2R_f)\Bigg\{
e_Q^{2}e_f^{2}\qquad \nonumber \\
+ \frac{2e_Q e_f v_Q v_f}{\sin^2\theta_W\cos^2\theta_W}\,\frac{1}{1-R_Z}
+ \frac{v_Q^{2}v_f^{2}}{\sin^4\theta_W\cos^4\theta_W}\,\frac{1}{(1-R_Z)^{2}}
\Bigg \}\qquad \nonumber \\
+ \beta_f^{2}\,\frac{v_Q^{2}a_f^{2}}{\sin^4\theta_W\cos^4\theta_W}\,\frac{1}{(1-R_Z)^{2}}
\Bigg]
|\psi(0)|^{2}
\end{align}
}
where, $\beta_f=\sqrt{1-4R_f}$ and, $v_i=\frac{1}{2}(I_{3L}+I_{3R})-2e_i\sin^2\theta_W$ and $a_i=\frac{1}{2}(I_{3L}-I_{3R})$ represent the vector and axial-vector couplings of the quark (Q) or fermion (f) to the $Z^0$ boson, respectively \cite{PhysRevD.35.3366}. Here, $(I_{3L})$ and $(I_{3R})$ denotes the third component of the weak isospin for left-handed and right-handed fermions respectively. In the Standard Model, all right-handed fermions are weak-isospin singlets with $(I=0)$ and $(I_3=0)$, whereas all left-handed fermions belong to weak-isospin doublets with $(I=\frac{1}{2})$ and $(I_3=\pm\frac{1}{2})$ \cite{baez2010algebragrandunifiedtheories}.

\begin{figure}[H]
\centering
\begin{tikzpicture}
\begin{feynman}

\vertex (a) at (-1,1) {$t$};
\vertex (b) at (-1,-1) {$\bar{t}$};
\vertex (c) at (0,0);

\vertex (z) at (1.5,0) {$Z^0$};

\vertex (f1) at (3,1) {$f$};
\vertex (f2) at (3,-1) {$\bar{f}$};

\diagram* {
(a) -- [fermion] (c) -- [fermion] (b),
(c) -- [boson] (z),
(z) -- [fermion] (f1),
(z) -- [anti fermion] (f2),
};

\end{feynman}
\end{tikzpicture}

\caption{\footnotesize $\eta_t \rightarrow f\bar{f}$ via $Z^0$ exchange}
\end{figure}
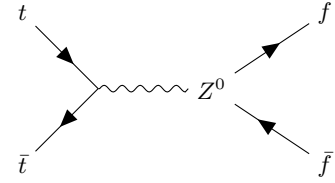

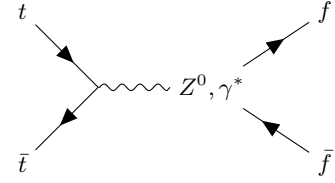
\begin{figure}[H]
\centering
\begin{tikzpicture}
\begin{feynman}
\vertex (a) at (-1,1) {$t$};
\vertex (b) at (-1,-1) {$\bar{t}$};
\vertex (c) at (0,0);

\vertex (z) at (1.5,0) {$Z^0,\gamma^*$};

\vertex (f1) at (3,1) {$f$};
\vertex (f2) at (3,-1) {$\bar{f}$};

\diagram* {
(a) -- [fermion] (c) -- [fermion] (b),
(c) -- [boson] (z),
(z) -- [fermion] (f1),
(z) -- [anti fermion] (f2),
};

\end{feynman}
\end{tikzpicture}

\caption{\footnotesize $\Theta_t \rightarrow f\bar{f}$ via $Z^0 \,\,\text{or}\,\, \gamma^*$ exchange}
\end{figure}
Here, $f \bar f$ represents any of the lepton pairs or quark flavours such as $\mu^+\mu^-$, $c\bar c$, $b \bar b$, etc.
\\
The dominant contribution to the fermionic decay of quarkonium arises from annihilation through a virtual photon. For relatively light quarkonium states, the contribution of a virtual $Z^0$ boson is highly suppressed and can be neglected. In contrast, for heavy quarkonium systems such as toponium, electroweak effects become important, and the virtual $Z^0$-boson channel contributes significantly to the decay amplitude \cite{Hern_ndez_Tom__2025}. The vector state $(1^{--})$ can couple to both a virtual photon and a virtual $Z^0$ boson through their vector interactions. In contrast, a pseudoscalar state $(0^{-+})$ cannot couple to a single photon and since the vector current cannot couple to a pseudoscalar $(0^{-+})$ state due to its quantum numbers, the vector contribution is forbidden. Consequently, the decay $\eta_t \rightarrow Z^0 \rightarrow f\bar{f}$ is mediated exclusively by the axial-vector part of the $Z^0$ interaction  \cite{PhysRevD.35.3366}. 
\subsubsection{Bosonic decay}
At higher energy scales, bosonic decay channels, especially those involving massive gauge bosons become increasingly relevant \cite{Kilian_2015,Bergstrom:1985hp}.  These bosonic decays of the pseudoscalar quarkonia are computed as \cite{Yang__2026,PhysRevD.35.3366,Bai_2026} 

\begin{equation}\small
    \Gamma(\eta_t \to Z^0 \gamma) = \frac{8\pi\alpha_{\mathrm{em}}\alpha_Z}{27M_{\eta_t}^2}
(3 - 8\sin^2\theta_W)^2
\lambda^{1/2}(1, R_Z, 0)| \psi(0)|^2,
 \end{equation}\\
\begin{align}
    \Gamma(\Theta_t\rightarrow Z^0\gamma)=\frac{8\pi\alpha_{em}\alpha_Z(1-R_Z^2)}{9M_{\Theta_t}^2R_Z}|\psi(0)|^2
\end{align}
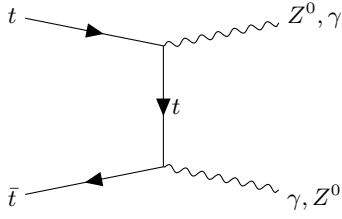
\begin{figure}[h]
\centering
\begin{tikzpicture}
\begin{feynman}
\vertex (i1) at (0,1.2) {\(t\)};
\vertex (i2) at (0,-1.2) {\(\bar t\)};
\vertex (v1) at (2,0.8);
\vertex (v2) at (2,-0.8);

\vertex (o1) at (4,1.2) {\(Z^0,\gamma\)};
\vertex (o2) at (4,-1.2) {\(\gamma,Z^0\)};

\diagram*{
(i1) -- [fermion] (v1)
      -- [fermion, edge label=\(t\)] (v2)
      -- [fermion] (i2),

(v1) -- [boson] (o1),
(v2) -- [photon] (o2),
};
\end{feynman}
\end{tikzpicture}
\caption{\footnotesize \(t\bar t \to Z^0\gamma\) through top exchange for both $\eta_t(0^{-+})$ and $\Theta_t(1^{--})$.}
\end{figure}

The decay process $t\bar t \rightarrow Z^0\gamma$ proceeds via t- and u-channel\footnote{The \(s\)-, \(t\)-, and \(u\)-channels denote different interaction topologies in Feynman diagrams, characterized by the Mandelstam variables \(s\), \(t\), and \(u\). The \(s\)-channel corresponds to particle annihilation via an intermediate state, while the \(t\)- and \(u\)-channels involve the exchange of a virtual particle.} top-quark exchange. For the pseudoscalar state $\eta_t(0^{-+})$, the decay is mediated through the vector component of the $Z^0t\bar t$ interaction. Whereas, the vector state $\psi_t(1^{--})$ couples through the axial-vector component \cite{PhysRevD.35.3366}. Unlike the $Z^0\gamma$ decay channel, which proceeds through top-quark exchange, the decay $t\bar t \rightarrow W^+W^-$ proceeds primarily through the exchange of the SU(2) partner quark in the t-channel, which would be the bottom quark for top quark, together with s-channel $\gamma^*$ and $Z^0$ contributions \cite{PhysRevD.35.3366}.
\begin{equation}
\Gamma(\eta_t \to W^+ W^-) = \frac{6\pi\alpha_W^2}{M_{\eta_{t}}^2}
\lambda^{-1/2}(1, R_W, R_W)|\psi(0)|^2, 
\end{equation}

{\small
\begin{align}
    \Gamma(\Theta_t\rightarrow W^+W^-)=\frac{\pi\alpha_{W}^2}{16M_\Theta^2R_W^2}\lambda^{3/2}(1,R_W,R_W)\qquad\nonumber\\
    \Bigg[\frac{1+20R_W+12R_W^2}{(1-R_Z)^2}\Bigg(1-\frac{16}{3}\sin^2\theta_WR_Z+\frac{64}{9}\sin^4\theta_WR_Z^2\Bigg)\quad\nonumber\\
    -\frac{8R_W(5+6R_W)(1-\frac{8}{3}\sin^2\theta_WR_Z)}{(1-4R_W)(1-R_Z)}+\frac{16R_W(2-R_W)}{(1-4R_W)^2}\Bigg]|\psi(0)|^2
\end{align}
}
\begin{figure}[h]
\centering
\begin{tikzpicture}
\begin{feynman}

\vertex (i1) at (0,1.2) {\(t\)};
\vertex (i2) at (0,-1.2) {\(\bar t\)};

\vertex (v1) at (2,0.8);
\vertex (v2) at (2,-0.8);

\vertex (o1) at (4,1.2) {\(W^+\)};
\vertex (o2) at (4,-1.2) {\(W^-\)};

\diagram*{
(i1) -- [fermion] (v1)
      -- [fermion, edge label=\(b\)] (v2)
      -- [fermion] (i2),

(v1) -- [boson] (o1),
(v2) -- [boson] (o2),
};

\end{feynman}
\end{tikzpicture}
\caption{\footnotesize \(\eta_t \rightarrow W^+W^-\) through t-channel bottom-quark exchange.}
\end{figure}
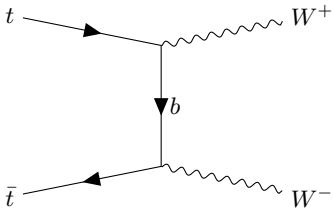

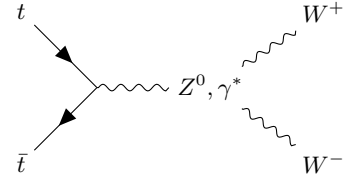
\begin{figure}[h]
\centering
\begin{tikzpicture}
\begin{feynman}

\vertex (a) at (-1,1) {$t$};
\vertex (b) at (-1,-1) {$\bar{t}$};
\vertex (c) at (0,0);

\vertex (z) at (1.5,0) {$Z^0,\gamma^*$};

\vertex (f1) at (3,1) {$W^+$};
\vertex (f2) at (3,-1) {$W^-$};

\diagram* {
(a) -- [fermion] (c) -- [fermion] (b),
(c) -- [boson] (z),
(z) -- [boson] (f1),
(z) -- [boson] (f2),
};

\end{feynman}
\end{tikzpicture}

\caption{\footnotesize $\Theta_t \rightarrow W^+W^-$ via $Z^0 \,\,\text{or}\,\, \gamma^*$ exchange}
\end{figure}
The $Z^0Z^0$ decay channel is mediated by t- and u-channel top-quark exchange, similar to the $Z^0\gamma$ decay process. Since the $Z^0$ boson couples directly to the top quark through the neutral-current interaction, the exchanged particle remains a top quark \cite{PhysRevD.35.3366}. The same mechanism is also observed in the $Z^0H$ decay channel.
{\small
\begin{equation}
    \Gamma(\eta_t \to Z^0 Z^0) = \frac{\pi\alpha_Z^2}{108M_{\eta_{t}}^2}
\frac{(9 - 24\sin^2\theta_W + 32\sin^4\theta_W)^2}{(1 - 2R_Z)^2
\lambda^{-3/2}(1, R_Z, R_Z)}|\psi(0)|^2, 
\end{equation}
}
{\small
\begin{align}
    \Gamma(\Theta\rightarrow Z^0Z^0)=\frac{\pi\alpha_{Z}^2(1-\frac{8}{3}\sin^2\theta_W)^2}{8M_\Theta^2(1-2R_Z)^2R_Z}\lambda^{5/2}(1,R_Z,R_Z)|\psi(0)|^2
\end{align}
}
\begin{figure}[h]
\centering
\begin{tikzpicture}
\begin{feynman}
\vertex (i1) at (0,1.2) {\(t\)};
\vertex (i2) at (0,-1.2) {\(\bar t\)};
\vertex (v1) at (2,0.8);
\vertex (v2) at (2,-0.8);

\vertex (o1) at (4,1.2) {\(Z^0\)};
\vertex (o2) at (4,-1.2) {\(Z^0\)};

\diagram*{
(i1) -- [fermion] (v1)
      -- [fermion, edge label=\(t\)] (v2)
      -- [fermion] (i2),

(v1) -- [boson] (o1),
(v2) -- [photon] (o2),
};
\end{feynman}
\end{tikzpicture}
\caption{\footnotesize \(t\bar t \to Z^0Z^0\) through top exchange for both $\eta_t(0^{-+})$ and $\Theta_t(1^{--})$.}
\end{figure}
The $Z^0H$ decay channel also proceeds through t- and u-channel top-quark exchange diagrams involving both the $Z^0t\bar t$ vertex and the Higgs-Yukawa coupling \cite{2014,PhysRevD.35.3366}. Unlike the $Z^0\gamma$ channel, the scalar nature of the Higgs boson changes the parity structure of the process, causing the pseudoscalar and vector toponium states to couple differently to the vector and axial-vector components of the $Z^0$ boson \cite{PhysRevD.35.3366}.
\begin{equation}
\Gamma(\eta_t \to Z^0 H) = \frac{3\pi\alpha_Z^2}{16M_{\eta_{t}}^2}
\lambda^{3/2}(1, R_Z, R_H)\frac{1}{R_Z^2}
|\psi(0)|^2.
\end{equation}
{\small
\begin{align}
    \Gamma(\Theta\rightarrow Z^0H)=\frac{\pi\alpha_{Z}^2(1-\frac{8}{3}\sin^2\theta_W)^2}{8M_{\Theta_t}^2}\lambda^{1/2}(1,R_H,R_Z)\nonumber\\
    \times\frac{[(1-R_Z)^2-R_H(1-3R_Z)]^2}{R_Z(1-R_Z)^2(1-R_Z-R_H)^2}+\nonumber\\\frac{\frac{1}{2}R_Z[(1-R_Z)^2+R_H(2-R_H)]^2}{R_Z(1-R_Z)^2(1-R_Z-R_H)^2}|\psi(0)|^2
\end{align}
}
\begin{figure}[h]
    \centering
\begin{tikzpicture}
\begin{feynman}
\vertex (t1) at (0,1.2) {\(t\)};
\vertex (tb1) at (0,-1.2) {\(\bar t\)};

\vertex (v1) at (2,0.8);
\vertex (v2) at (2,-0.8);

\vertex (z) at (4,1.2) {\(Z,H\)};
\vertex (h) at (4,-1.2) {\(H,Z\)};

\diagram*{
(t1) -- [fermion] (v1)
      -- [fermion, edge label=\(t\)] (v2)
      -- [fermion] (tb1),

(v1) -- [boson] (z),
(v2) -- [scalar] (h),
};
\end{feynman}
\end{tikzpicture}
\caption{\footnotesize \(t\bar t \to Z^0H\) through top exchange for both $\eta_t(0^{-+})$ and $\Theta_t(1^{--})$.}
\end{figure}
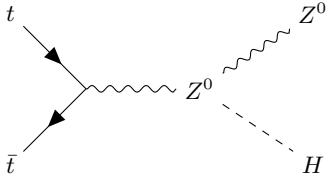
\begin{figure}[h]
\centering
\begin{tikzpicture}
\begin{feynman}

\vertex (a) at (-1,1) {$t$};
\vertex (b) at (-1,-1) {$\bar{t}$};
\vertex (c) at (0,0);

\vertex (z) at (1.5,0) {$Z^0$};

\vertex (f1) at (3,1) {$Z^0$};
\vertex (f2) at (3,-1) {$H$};

\diagram* {
(a) -- [fermion] (c) -- [fermion] (b),
(c) -- [boson] (z),
(z) -- [boson] (f1),
(z) -- [scalar] (f2),
};

\end{feynman}
\end{tikzpicture}
\caption{\footnotesize $\Theta_t \rightarrow Z^0H$ via $Z^0$ exchange}
\end{figure}

The numerical values of particle masses and fundamental physical constants used in this work are taken from \cite{PhysRevD.110.030001}. The leptons includes the tau mass $m_\tau = 1776.93\,\text{MeV}$ and the muon mass $m_\mu = 105.658\,\text{MeV}$. For quarks, the top quark mass is $m_t = 172.52\,\text{GeV}$, the bottom quark mass is $m_b = 4.78\,\text{GeV}$, and the charm quark mass is $m_c = 1.67\,\text{GeV}$. For the bosonic sector, the Higgs boson mass is $m_H = 125.20\,\text{GeV}$, the $Z^0$ boson mass is $m_Z = 91.1880\,\text{GeV}$, and the $W$ boson mass is $m_W = 80.3692\,\text{GeV}$ \cite{PhysRevD.110.030001}. The relevant physical constants include the electromagnetic coupling constant, $\alpha_{\text{em}}(m_W) = 1/128$, the strong coupling constant at the $Z^0$ boson scale, $\alpha_s(m_Z) = 0.11$, and the weak mixing parameter $\sin^2\theta_W = 0.23129$ \cite{PhysRevD.110.030001}. Making use of the radial wave function of the bound state and the masses of the relevant states, the various decay widths are computed for the $\eta_t$ and $\Theta_t$ states and are tabulated in tables \ref{tab:placeholder2} and \ref{tab:placeholder3} respectively.

\begin{table}[H]
    \centering
    \setlength{\tabcolsep}{9.7pt}
    \begin{tabular}{lccc}
    \hline 
       Channel &  Width [Our] & Ref.\cite{Yang__2026} & Ref.\cite{fu2025toponiumimplementationtoponiummodel}\\
        \hline
        \hline
        $gg$ & 2.57 MeV & 1.99 MeV &1.59 MeV  \\
        $\gamma \gamma$  & 9.88 keV & 9.32 keV & 9.20 keV\\
        $b\bar b$ & 1.49 keV & 1.36 keV & -\\
        $c\bar c$ & 0.18 keV  & 0.17 keV & -\\
        $\mu^+\mu^-$ & 0.24 eV & 0.22 eV&-\\
        $\tau^+\tau^-$ & 68.46 eV & 62.68 eV&-\\
        $Z^0 \gamma$ & 2.19 keV & 2.01 keV& 2.30 keV\\
        $W^+W^-$ & 135.39 keV & 97.45 keV& 130.20 keV\\
        $Z^0Z^0$ & 6.87 keV & 6.32 keV& 6.70 keV\\
        $Z^0H$ & 594.68 keV & 537.43 keV & 562.40 keV\\
    \hline
    \end{tabular}
    \caption{Pseudoscalar ($0^{-+}$) state decay widths of toponium calculated within the non-relativistic SUSY QM framework using the Cornell potential.}
    \label{tab:placeholder2}
\end{table}
\begin{table}[H]
    \centering
    \setlength{\tabcolsep}{5.5pt}
    \begin{tabular}{lccc}
    \hline
        Channel & Width [Our] & Ref.\cite{fu2025toponiumimplementationtoponiummodel} & Ref.\cite{jiang2024studytoponiumspectrumassociated}\\
        \hline
        \hline
          $ggg$ &  57.72 eV & - & 52.70 eV\\
        $\gamma \gamma \gamma$ & 13.93 eV & - & 15.00 eV\\
        $b\bar b$ & 454.98 keV & 491.40 keV & -\\
        $c\bar c$ & 12.87 keV & 13.40 keV & - \\
        $e^+e^-,\mu^+\mu^-,\tau^+\tau^-$ & 7.36 keV & 7.80 keV & -\\
        $\nu_e\bar\nu_e,\nu_\mu\bar\nu_\mu,\nu_\tau\bar\nu_\tau$ & 0.38 keV & 0.80 keV & -\\
      
        $Z^0\gamma$ & 75.57 keV & 70.30 keV & -\\
        $W^+W^-$ & 570.45 keV & 546.60 keV & -\\
        $Z^0Z^0$ & 5.24 keV & 5.50 keV & -\\
        $Z^0H$ & 7.71 keV & 8.10 keV & -\\
    \hline
    \end{tabular}
    \caption{Vector ($1^{--}$) state decay widths of toponium calculated within the non-relativistic SUSY QM framework using the Cornell potential.}
    \label{tab:placeholder3}
\end{table}

In the framework of the Standard Model, the leading-order (LO) decay width of the top quark can be expressed through the Fermi constant 
as \cite{Gao_2013}:
\begin{equation}
\small
\Gamma_t^{(0)} = \frac{G_F\, m_t^3}{8\sqrt{2}\,\pi}
\left[1 - 3\left(\frac{m_W^2}{m_t^2}\right)^2 + 2\left(\frac{m_W^2}{m_t^2}\right)^3 \right]
\end{equation}
This expression is obtained under the simplifying assumptions that the relevant CKM matrix element satisfies $|V_{tb}| = 1$, and that the bottom quark mass is negligible ($m_b = 0$). Physically, this corresponds to the dominant decay channel $t \rightarrow W b$, where phase-space effects and the finite mass of the $W$ boson are fully taken into account. Using the leading-order expression for the total top quark decay width, the numerical value is obtained as $\mathbf{\Gamma_t = 1.4783 \,\text{GeV}}$. Taking $\hbar = 6.5821 \times 10^{-25} \,\text{GeV}\cdot\text{s}$, the corresponding top quark lifetime is found to be $\mathbf{\tau_t = 4.452 \times 10^{-25} \,\text{s}}$. 

For the toponium ($t\bar{t}$) system, and under the assumption that the constituent top quark and  the anti-top quark decay independently \cite{Zhang_2025}, the total decay width can be approximated as $\mathbf{\Gamma_{t\bar{t}} = 2\Gamma_t = 2.9566 \,\text{GeV}}$. Consequently, the lifetime of the toponium system is reduced to $\mathbf{\tau_{t\bar{t}} = 2.226 \times 10^{-25} \,\text{s}}$.

\section{Results and Discussion}

The application of the Supersymmetric Quantum Mechanics (SUSY QM) formalism to the Cornell potential demonstrates that the method can be consistently extended to physically relevant heavy quark systems such as toponium. The computed mass spectrum, binding energy, and hyperfine splitting collectively provide a coherent and physically intuitive description of heavy quark dynamics, particularly in the short-distance regime. The binding energy of approximately $E_B \approx -0.9056\,\text{GeV}$, indicates the formation of a tightly bound state. The corresponding masses of the pseudoscalar and vector states are estimated to be around $344.114\,\text{GeV}$ and $344.141\,\text{GeV}$, respectively, which are in good agreement with the previously predicted mass results reported in \cite{Najjar2026,https://doi.org/10.1155/ahep/9454944,PhysRevD.111.096016,jiang2024studytoponiumspectrumassociated}, while the hyperfine splitting between these states is found to be $27.2\,\text{MeV}$. These results are consistent with the expected behavior of a highly compact heavy quarkonium system dominated by short-range interactions. In addition, the binding energy provides insight into possible dissociation in extreme environments such as a quark–gluon plasma, linking collider physics with early-universe conditions \cite{benzahra2005bindingenergiesdissociationsjpsi}.

Within this framework, both pseudoscalar and vector channels exhibit well-defined bound-state characteristics governed by the wavefunction at the origin $|\Psi(0)|^2$, which directly controls the decay properties. Unlike charmonium and bottomonium, the decay pattern of toponium is qualitatively different. Electromagnetic channels such as $\eta_t\rightarrow\gamma\gamma$ and $\Theta_t\rightarrow \ell^+\ell^-$ are strongly suppressed with decay widths of 9.88 keV and 7.36 keV, respectively, in agreement with the results presented in \cite{Yang__2026,fu2025toponiumimplementationtoponiummodel}, while weak decay modes become dominant due to the large top-quark mass. The $\eta_t \to gg$ channel remains the leading QCD contribution with the decay width of 2.55 MeV, though experimentally challenging due to strong hadronic backgrounds, whereas cleaner signatures may arise from photonic and electroweak bosonic channels \cite{Yang__2026}. The calculated decay widths for the bosonic and fermionic channels of toponium reveal a clear hierarchy between the two classes of decays. Among the electroweak bosonic channels, the $\eta_t\rightarrow Z^0H$ and $\Theta_t\rightarrow W^+W^-$ mode yield the largest electroweak decay width of $594.68~\text{keV}$ and $570.45~\text{keV}$, followed by the $\eta_t\rightarrow W^+W^-$ and $\Theta_t\rightarrow Z^0\gamma$ channel with a decay width of $135.39~\text{keV}$ and $75.57~\text{keV}$, respectively, consistent with the decay hierarchy discussed previously in \cite{Yang__2026,Bai_2026}. In this work, the combined use of the Cornell potential and SUSY QM techniques ensures that both the long-range confining behavior and the short-distance Coulombic interactions are properly incorporated, thereby providing a consistent framework for analyzing such high-energy decay processes. In contrast, the fermionic decay channels $\eta_t \rightarrow f\bar f$ are found to be strongly suppressed, consistent with the results presented in \cite{Yang__2026}, with the estimated decay widths of $1.49~\text{keV}$ and $0.18~\text{keV}$ for the quark channels $b\bar b$ and $c\bar c$, respectively, and $0.24~\text{eV}$ and $68.46~\text{eV}$ for the leptonic channels $\mu^+\mu^-$ and $\tau^+\tau^-$, respectively. The resulting event estimated for all fermionic channels are found to be negligibly small under the projected HL-LHC luminosity conditions \cite{Fuks_2021,book}, rendering these modes effectively inaccessible to current experimental sensitivities and confirming that the bosonic decay channels constitute the primary experimental handles for the discovery and characterization of the pseudoscalar toponium meson. In contrast, the fermionic decay channels of the vector state are significantly enhanced since they proceed through both $\gamma^*$ and $Z^0$ exchange, allowing contributions from the vector and axial-vector couplings, unlike the pseudoscalar state, which decays only through the $Z^0$ boson. Consequently, the $b\bar b$ channel is found to be one of the dominant decay modes as predicted by \cite{Bai_2026}, with an estimated partial decay width of $454.98~\text{keV}$.

The study of spatial observables such as the mean radius, RMS radius, and most probable radius further strengthens the connection between the SUSY QM formalism and the physical structure of the bound state. These quantities also play an important role in collider phenomenology, where accurate modeling of the compact system aids in distinguishing signals from standard $t\bar{t}$ production \cite{Kats_2010}. From a broader perspective, toponium provides a unique testing ground for Quantum Chromodynamics. Its extremely small spatial extent with a most probable radius of 0.015 fm places it deep within the perturbative regime, where short-range gluon exchange dominates and allows for direct tests of pQCD predictions \cite{Top_BS}. The hyperfine splitting and binding energy are particularly sensitive to the interquark potential, offering a way to define precise quantities such as the top quark 1S mass. At the same time, the large top-quark mass enhances sensitivity to electroweak effects, making toponium a useful probe of Higgs-top Yukawa couplings and possible physics beyond the Standard Model \cite{2014}.

Several extensions could further improve the present analysis. Incorporating the higher-order corrections to the potential and extending the study to excited states would provide a more complete and precise description of toponium-like systems.
\acknowledgments{The authors gratefully acknowledge the facilities and academic support provided by the P. D. Patel Institute of Applied Sciences (PDPIAS), CHARUSAT, Anand, Gujarat, India.}

\nocite{*}
\bibliographystyle{apsrev4-2}
\bibliography{References}

\end{document}